\documentclass{article}
\usepackage{spconf,amsmath,graphicx,subfig}

\title{CHAPTER: Exploiting Convolutional Neural Network Adapters for Self-supervised Speech Models}
\name{Author(s) Name(s)}
\address{Author Affiliation(s)}
\name{\centering Zih-Ching Chen, Yu-Shun Sung,  Hung-yi Lee}

\address{Graduate Institute of Communication Engineering, National Taiwan University, Taiwan}

\begin{document}
\ninept
\maketitle
\begin{abstract}
Self-supervised learning (SSL) is a powerful technique for learning representations from unlabeled data. Transformer based models such as HuBERT, which consist a feature extractor and transformer layers, are leading the field in the speech domain. SSL models are fine-tuned on a wide range of downstream tasks, which involves re-training the majority of the model for each task. Previous studies have introduced applying adapters, which are small lightweight modules commonly used in Natural Language Processing (NLP) to adapt pre-trained models to new tasks. However, such efficient tuning techniques only provide adaptation at the transformer layer, but failed to perform adaptation at the feature extractor. In this paper, we propose CHAPTER, an efficient tuning method specifically designed for SSL speech model, by applying CNN adapters at the feature extractor. Using this method, we can only fine-tune fewer than 5\% of parameters per task compared to fully fine-tuning and achieve better and more stable performance. We empirically found that adding CNN adapters to the feature extractor can help the adaptation on emotion and speaker tasks. For instance, the accuracy of SID is improved from 87.71 to 91.56, and the accuracy of ER is improved by 5\%. 

\end{abstract}
\begin{keywords}
Efficient tuning, Adapter, Self-supervised models, Speech processing, SUPERB
\end{keywords}

\section{Introduction}
\label{sec:intro}
\begin{figure}

      \centering
      \includegraphics[width=0.4\textwidth]{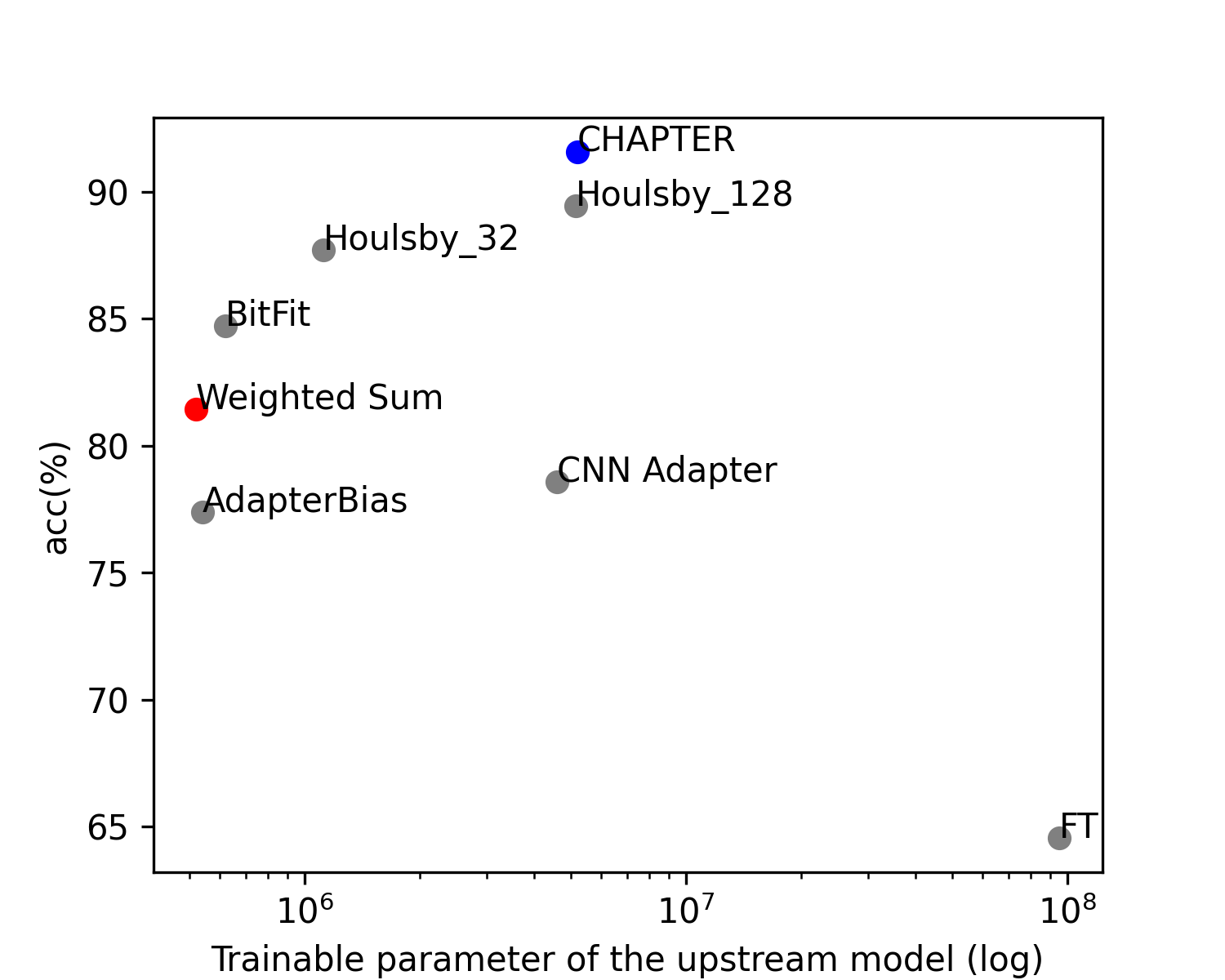}
      \caption{The trade-off between accuracy and trainable parameters. The x-axis represents trainable parameter of the upstream model, while the y-axis represents the accuracy of Speaker Identification task (SID). The red point is our baseline, weighted sum, which is reported in the SUPERB benchmark, and the blue point is our proposed CHAPTER. We also compared CHAPTER to different PET methods in NLP \cite{houlsby2019parameter, zaken2021bitfit, fu-etal-2022-adapterbias}.
      }
\label{fg:1_overview}
\end{figure}
Self-supervised learning has gained huge popularity in the field of computer vision (CV), natural language processing, as well as speech processing tasks. SSL pre-trains a shared representation model on a huge amount of unlabeled data. Applying a SSL model for various downstream tasks with adaptation can significantly lower the entry barrier for developing a model compared to training the model from scratch\cite{Mohamed2022SSLspeech}.

Despite the huge success and popularity SSL has gained, some drawbacks exist when utilizing SSL models. In the presence of various downstream tasks, fine-tuning pre-trained models for each downstream task is parameter-inefficient, as one copy of model parameter has to be kept for each task in memory and storage, plus SSL models often have millions or billions of parameters. Due to this reason, utilizing the SSL speech model by fine-tuning requires large storage space. 
\begin{figure*}
\vspace{-5pt}
\centering
      \includegraphics[width=0.95\textwidth]{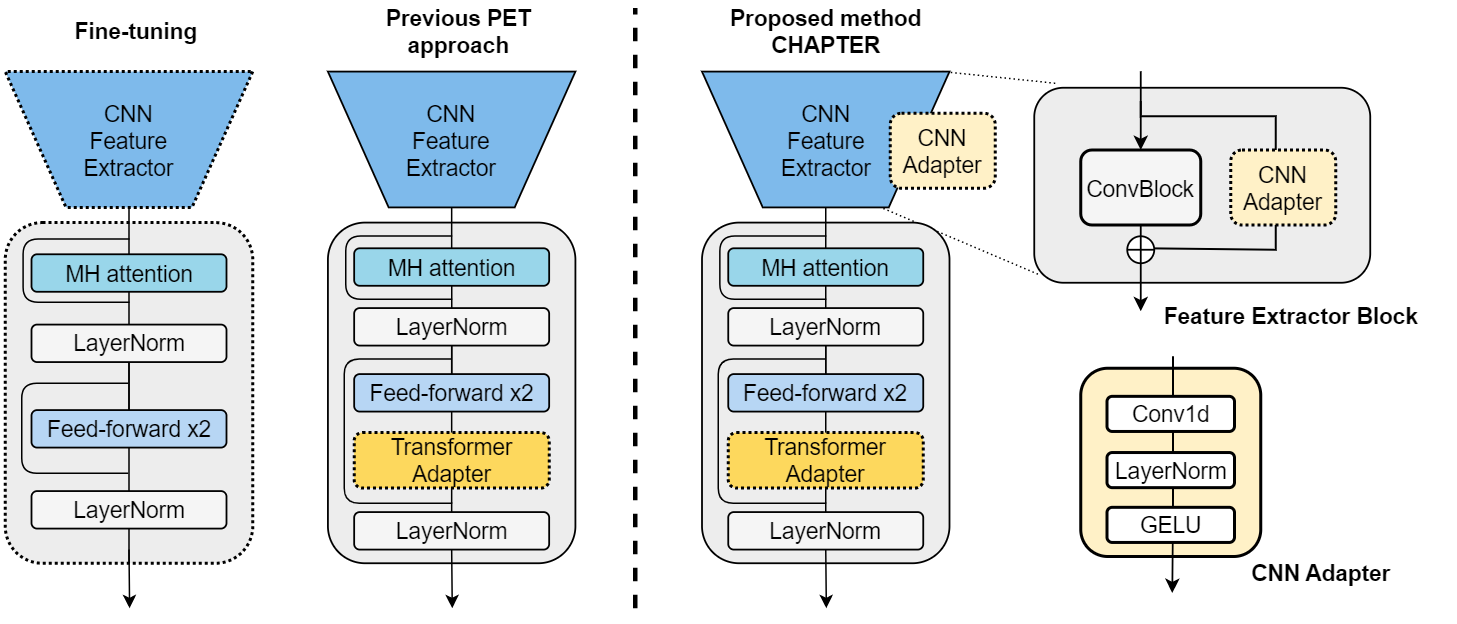}
\caption{
        Illustration of our overall framework compared to previous approach. Previous approaches focused on exploring transformer-based adapters, while our proposed method adds CNN adapter modules to the feature extractor, which is specifically designed for PET of SSL speech models.
      }

\label{fg:overall_architecture}
\end{figure*}

To overcome these shortcomings, researchers then utilize the SSL speech model by only using the frozen representation~\cite{yang2021superb}. In NLP, parameter efficient tuning (PET) techniques have been proposed for leveraging SSL models. One of the most popular efficient methods is adapters~\cite{houlsby2019parameter}, which introduce extra tunable weights and freeze the original parameters of the pre-trained language model (PLM). Adapters have demonstrated comparable performance with fully fine-tuning the entire model while being parameter-efficient. Rather than requiring a separate copy of the model for each downstream task, a single generalized upstream model can simultaneously transfer to many different tasks.

In speech processing tasks, adapters have also been utilized for efficient SSL tuning. Using adapters on Wav2vec2\cite{baevski2020wav2vec} for efficient tuning for ASR has been proposed~\cite{thomas2022efficient}.
SimAdapter~\cite{hou2021exploiting} explicitly learns knowledge from adapters, which has very similar concept with \cite{pfeiffer2020adapterfusion}. 
Adapters have also been employed for efficient SSL speech pre-training of new tasks in a continual learning setting~\cite{kessler2022adapter}. More recently, \cite{chen2022exploring,yang2021voice2series} proposed a unified framework for adapters applications to a wide range of speech processing tasks. 

Although adapters have been applied to SSL speech model, these PET approaches only directly apply NLP adapters to SSL speech models, without exploring the design of efficient tuning approaches based on the attributes of audio signals. A convolutional neural network (CNN) is a typical component in Audio/speech SSL models to learn from spectrogram and tokenize audio input. CNN is critical in bridging raw audio input and tokens where Transformer learns contextual representation from. A typical SSL speech model consists of a feature extractor, which takes raw audio as input and outputs latent speech representations, and transformer layers, while NLP models only have transformer layers. Existing adapters are mostly designed based on NLP model, which is different from the typical SSL speech model consisting of a feature extractor and transformers. 
In CV, \cite{rebuffi2017learning} and \cite{rebuffi2018efficient} proposed using residual adapters in CNN in order to learn multiple visual domains. However, there is no work exploring adaptation of the CNN component. 

Based on this view, we propose CHAPTER, a PET method with CNN adapters on the CNN feature extractor in front of the SSL speech model for better adaptation and transformer adapters to the transformer module. CNN adapters can better adapt the feature extractor model while the transformer adapter learns how to utilize the representation from the transformer layers much more efficiently. Experiments have been conducted on the SUPERB benchmark~\cite{yang2021superb} to investigate the effectiveness of our proposed CHAPTER on HuBERT \cite{hsu2021hubert}. We found that CNN adapters work especially well on tasks relying more on the feature extractor, including speaker related and emotion tasks. We show that the performance parity can be achieved with over 90\% parameter reduction. Compared to applying Houlsby adapter only, our proposed CHAPTER adapter can improve the accuracy of Speaker Identification (SID) from 87.71\% to 91.56\%, as shown in Fig~\ref{fg:1_overview}; and improve Emotion Recognition task (ER) from 65.25\% to 70.41\%.

\section{Methods}
\subsection{Preliminaries}
PET methods introduce a set of additional learnable modules plugged into the backbone, which is frozen during training. This efficient adaptation module aims to learn a task-specific adaptation from the original hidden representation to the downstream task.
To be more specific, considering an intermediate hidden representation ${h}$ generated by a layer
or a series of layers with input ${x}$ in a pre-trained network, the adapter module learns ${\Delta h}$ and updates ${h}$ as:
\begin{equation}
    h \leftarrow h + \alpha \cdot \Delta h
\end{equation}
where ${\alpha}$ represents a scalar or a gating function. Previous adapter methods in NLP followed a similar functional form for construction ${\Delta h}$, including downsampling projection, non-linearity, and upsampling projection \cite{he2022towards}. In CV, \cite{rebuffi2017learning} proposed adding residual adapters to ResNet in order to adapt to multiple domains.
Previous PET mainly focused on adding transformer adapters to an SSL speech model. In our work, we construct ${\Delta h}$ in the feature extractor part with CNN adapter; and ${\Delta h}$ of the transformer layer is constructed with Houlsby adapter.

For our experiments, we applied different downstream models, including LSTM module or a linear classifier, on top of the transformer network following the setting of the SUPERB benchmark. A set of adapters, including CNN adapters and Houlsby adapters, and the downstream model were trained per task, and the rest of the network remained frozen.
\subsection{CHAPTER}
\begin{table*}
\renewcommand{\arraystretch}{1.2}
\centering
\setlength{\tabcolsep}{3.5mm}{
\begin{tabular}{c|c|ccccccccc}
\hline
Method         & Params         & ASR           & PR            & SF             & SID            & SD            & SV            & IC             & KS             & ER             \\ \hline
FT             & 94.7M          & 6.35 & \textbf{2.45} & 86.17          & 64.56          & \textbf{3.58}         & 5.15             & 99.1           & 95.81          & 68.94          \\
Baseline       & 0              & 7.09          & 7.74          & 81.67          & 64.78          & 7.05          & 5.31          & 96.39          & 95.32          & 65.16          \\
Houlsby$_{32}$        & 0.60M          & \textbf{5.88}          & 3.00          & 84.68          & 87.71          & 4             & 5.29 & \textbf{99.6}  & 97.17          & 65.25          \\
Weighted-sum     & 12             & 6.42          & 5.41         & \textbf{86.71} & 81.42          & 5.88          & 5.11          & 98.34          & 96.3           & 64.92          \\
CNN adapter         & 4.07M          & 6.32          & 5.42          & 86.81          & 78.57          & 5.7           & 5.41             & 98.39 & \textbf{97.2}  & 64.147         \\
CHAPTER & 4.67M & 6.22 & 2.95          & 85.94                 & \textbf{91.56} & 3.84 & \textbf{4.95} & 99.2           & 95.52 & \textbf{70.41} \\ \hline
\end{tabular}

}

\caption{Performance of different efficient methods in the SUPERB benchmark. The second column represents additional trainable parameter used in upstream model. Note that in emotion recognition task (ER), we report the result of fold1 test.}
\label{tab:1}
\end{table*}
\subsubsection{CNN adapters} 

In this paper, we adapted the feature extractor of the SSL speech model by inserting CNN adapter modules to its CNN blocks. A CNN adapter consists of one Conv1d layer, a LayerNorm layer, followed by a GELU activation function.
Given the input $x_{in}$, $x_{in}$ pass a temporal convolution follow by layer normalization and a GELU activation function. As shown in Fig. \ref{fg:overall_architecture}, the feature extractor block would output the sum of CNN adapter module's output $x_{out}$ and the output of ConvBlock.

\begin{figure} 
  \centering
  \includegraphics[width=0.4\textwidth]{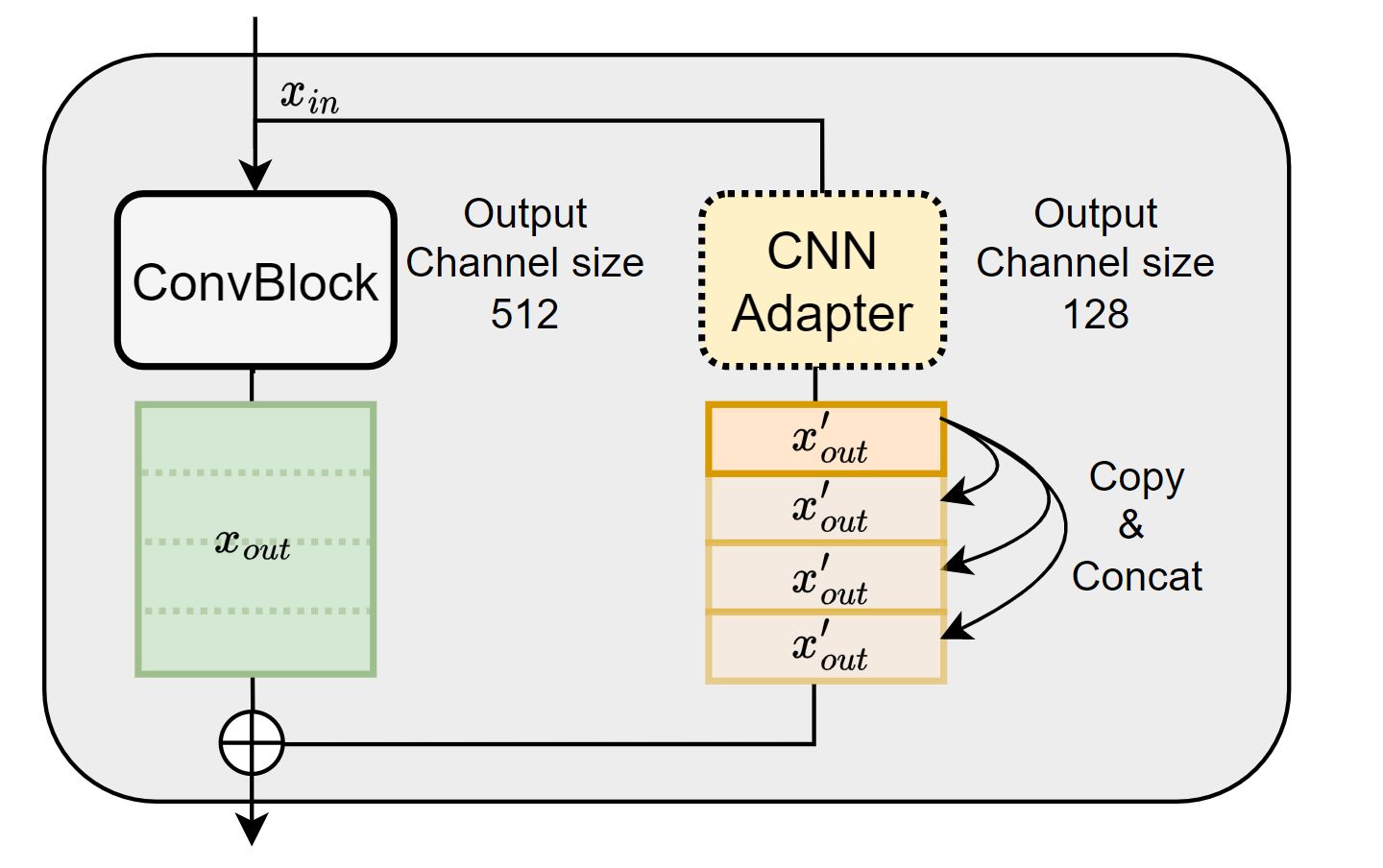}
  \caption{CNN Adapter compression. Note that the diagram shows the scenario of output channel compressed by 4 times ($n=4)$ and the original number of output channels is 512}
  \label{CNN compress}
\end{figure}

The motivation of adapter techniques is to improve parameter efficiency.To further explore this property on speech tasks, we decreased the parameters of our CNN adapter in two ways: shrinking each adapter layer and selectively-added adapter.
Shrinking each adapter layer was implemented in CNN adapter by decreasing the number of its output channels since the parameter number is proportional to it.
Afterward, we concatenated the $n$ copies of adapter layer's output  together on channel dimension, where $n$ denotes the ratio between original channel size and the compressed one. By doing so, the the concatenated output  
can match the shape of feature extraction layer's output for the adding operation later (see Fig. \ref{CNN compress}). As for selective-added adapter, it was implemented by only adding CNN adapter to some layers in feature extractor. Note that for the layers without adapter added, the parameters in original feature extraction layer were still frozen.

\subsubsection{Overview of CHAPTER}
In this paper, we propose CHAPTER, which adds CNN adapter to the feature extractor while inserting Houlsby adapters to the transformer layers. The overall architecture of CHAPTER is shown in Fig~\ref{fg:overall_architecture}.
We inserted Houlsby adapters to the transformer layers in SSL speech models for better modeling contextual information in downstream tasks. (see Fig.\ref{fg:overall_architecture}).
Houlsby adapters \cite{houlsby2019parameter} are small bottleneck modules consisting of a down-projection ($FF_{down}$), a non-linearity ($GeLU$), and an up-projection ($FF_{up}$), with a skip connection. Houlsby adapters are typically applied after both the self-attention and feed-forward layers \cite{houlsby2019parameter}.
In our work, Houlsby adapters\cite{houlsby2019parameter} were added to the second feed-forward layers of transformer layers, where the fully connected layers are initialized as a near identity function. The bottleneck of each inserted Houlsby adapter is set to 32.
\section{Experiment}
\label{sec:exp}

\subsection{Experiment setup}
To demonstrate the effectiveness of our proposed CHAPTER, we conducted experiments on the SUPERB benchmark\cite{yang2021superb}, which is a framework to benchmark SSL models on 10 speech tasks by learning task-specific predictions heads on top of the frozen shared SSL models. All the experiments strictly followed the the SSL speech parameter-efficient framework proposed by \cite{chen2022exploring}.
For the learning rate, we search for the best learning rate across $10^{-3}$ to $10^{-5}$ for each combination of SSL representation and the downstream tasks. 

\subsection{Performance on the SUPERB benchmark}
\label{exp:1}
To examine the effectiveness of our proposed CHAPTER, we compared it to different efficient methods in the SUPERB benchmark and the result is shown in Table~\ref{tab:1}. Note that 'FT' represents fine-tuning, 'CNN adapter' means that we only added CNN adapter to the feature extractor without utilizing Houlsby adapter in the transformer layers, and 'Houlsby$_32$' is Houlsby adapter with bottleneck 32. The 'Baseline' here means that we only tuned the downstream model. The tasks we have examined can be categorized into four: recognition task, classification task, speaker task, and emotion task. We found that CHAPTER performed very well on speaker related tasks, including speaker identification (SID), speaker verification(SV), as well as speaker diarization(SD) and emotion tasks (ER). Especially in SID, our proposed CHAPTER significantly reduces the error rate by 3.85\% compared to using Houlsby only, which is already a powerful approach with 87.71\% accuracy. In emotion tasks, CHAPTER improves the accuracy by 5.16\% compared to the original Houlsby adapter. Overall, CHAPTER yields outstanding performance on speaker and emotion tasks.

\begin{table}
\centering
\renewcommand{\arraystretch}{1.12}
\begin{tabular}{c|c|ccc}
\hline\hline
Method        & param(M)      & SID            & SD            & ER             \\[2pt] \hline
FT            & 94.7          & 64.56          & \textbf{3.58}          & 68.94          \\[2pt]
Houlsby$_{32}$       & 0.6           & 87.71          & 4             & 65.25          \\[2pt]
Houlsby$_{128}$ & 4.61          & 89.44          & 3.91          & 66.82          \\[2pt] \hline
CHAPTER & \textbf{4.67} & \textbf{91.56} & 3.84 & \textbf{70.41} \\ \hline\hline
\end{tabular}

\caption{Compared with larger Houlsby adapter. We examine the placement of additional parameters on ER and SD by comparing CHAPTER to Houlsby adapter with bottleneck size 128. For ER, here we present the accuracy of fold 1 testing set.}
\label{tab:compare_houlsby}
\end{table}

\subsection{Compared with Large Houlsby adapter}
On average, CHAPTER yields good performance among all efficient methods, espeically on the speaker related tasks and emotion task. Since CHAPTER has more trainable parameters compared to Houlsby$_{32}$, we want to know if such performance gain is only the result of more number of trainable parameters. To see if the improvement is because of adding CNN adapters to the feature extractor, we increased the bottleneck of Houlsby adapters from 32 to 128, so that its number of trainable parameters is comparable with CHAPTER. As shown in Table \ref{tab:compare_houlsby}, CHAPTER significantly outperforms Houlsby adapter$_{128}$. This shows that adding CNN adapters to the feature extractor can improves the performance on speaker and emotion tasks.

\begin{figure}
    \centering
      \includegraphics[width=0.45\textwidth]{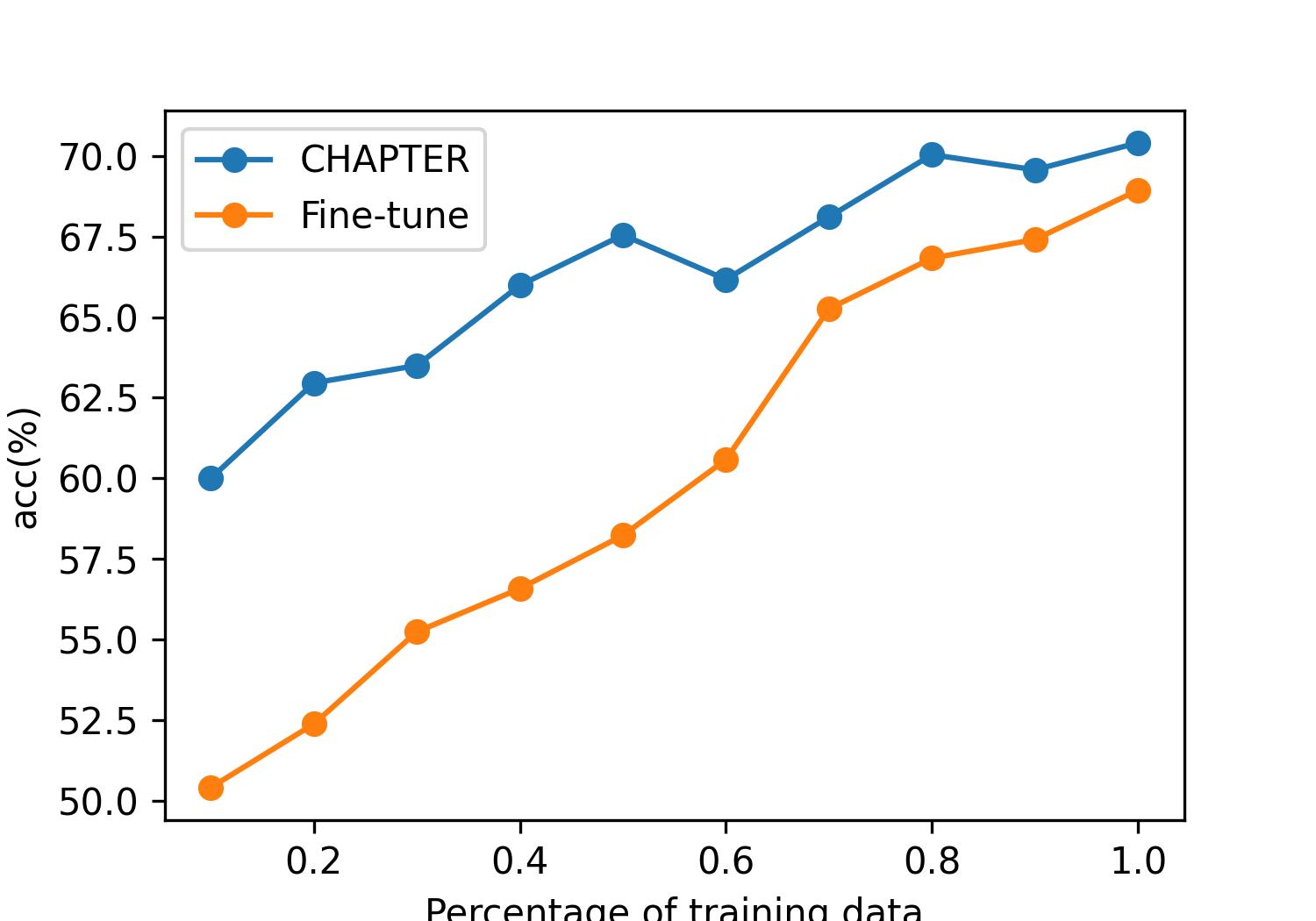}
      \caption{
        Size of small training data. The experiment is conducted on ER.
      }
\label{fg:small data}
\end{figure}

\subsection{Low-resource Adaptation}
In NLP, adapters have been shown to have advantages over fine-tuning when adapting to low-resource datasets \cite{he2021effectiveness, zaken2021bitfit, fu-etal-2022-adapterbias}. To check if this property also holds in CHAPTER when applied in speech tasks, we compare our CHAPTER with fine-tuning in the low-resource settings in the ER. We observed that CHAPTER less performance drop on low-resource datasets.  As the training data becomes smaller, tuning the majority of the parameters may result in a higher risk of overfitting the training data. Using CHAPTER method helps overcome this issue. As shown in Fig~\ref{fg:small data}, CHAPTER strictly outperforms fine-tuning.
\subsection{Ablation study}
\subsubsection{Top  layers}

\begin{table}
\centering
\setlength{\tabcolsep}{0.8mm}{
\begin{tabular}{c|c|cc}
\hline
Methods  & Params & ER (acc)       & SD (der)      \\[1pt] \hline 
Full     & 4.07M      & 70.41          & \textbf{3.58} \\[1pt]
Top5    & 3.4M       & 71.05          & 3.82          \\[1pt]
Top4    & 2.62M      & 69.48          & 3.82          \\[1pt]
Top3    & 1.83M      & \textbf{71.89} & 3.88          \\[1pt]
Top2    & 1.04M      & 70.13          & 3.96          \\[1pt]
Top1    & 0.524M     & 65.8           & 4.04          \\[1pt] \hline
\end{tabular}
\caption{Ablation study on adding CNN adapters at top N layers. Here, Top 5 means that we added CNN adapter on the top 5 layers. Also, Houlsby adapters were inserted in the transformer layers as well.}

\label{tab:last_N_layers}
}
\end{table}



\begin{table}
\centering
\begin{tabular}{c|c|cc}

\hline
Method            & param(M)        &SD           & ER             \\ 
\hline
FT                        & 94.7            &3.58         & 63.94          \\
\text {CHAPTER$_{512}$}                    & 4.07            &3.84         & 65.44          \\

\text {CHAPTER$_{256}$}                         & 2.10            &3.99         & 65.16          \\

\text {CHAPTER$_{128}$}                       & 1.05            &3.87         & 65.75          \\

\text {CHAPTER$_{64}$}                        & 0.525           &4.06         & 64.58          \\
 \hline 

\end{tabular}

\caption{Ablation study with CNN adapter layer compression. We tested on 5 fold cross validation in ER and the average score was reported. Here we reduced the output channel to make the CNN adapters more parameter efficient, where 512 is the original output channel size.}
\label{tab:CNN shrink}
\end{table}
\begin{figure}[h]
      \centering
      \includegraphics[width=0.45
      \textwidth]{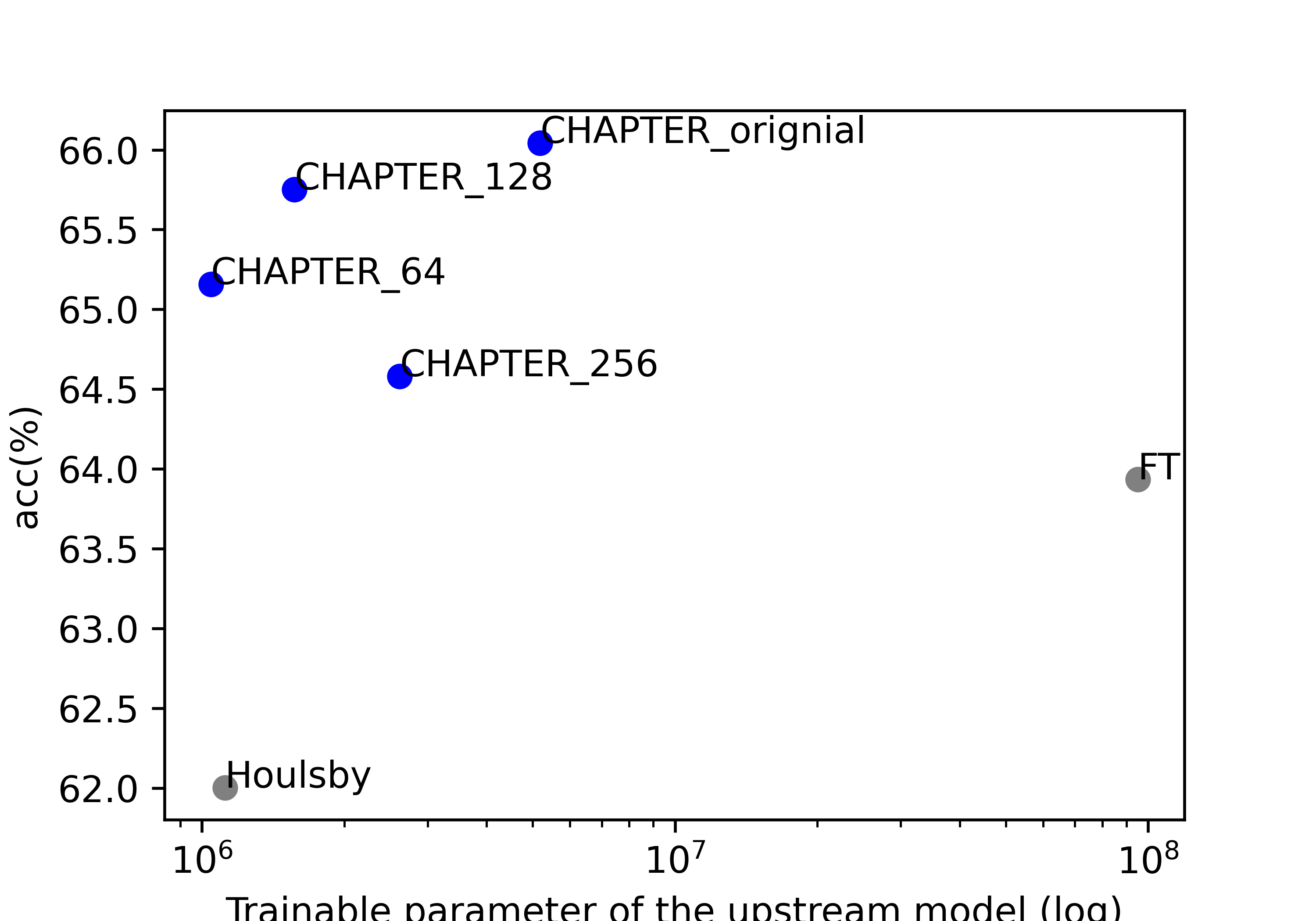}
      \caption{The trade-off between accuracy and trainable parameters. The x-axis represents trainable parameter of the upstream model, while the y-axis represents the accuracy of Speaker Identification task (ER). 
      }
\label{fg:overview}
\end{figure}
Prior work suggests that lower layers of self-supervised models contain more generic speech features, and higher layers contribute more to phone discrimination \cite{chung2019unsupervised}. To explore this trade-off, we consider different adapter sizes, and compare to adding only adapters to the top N layers of the feature extractor in HuBERT. As shown in Table.\ref{tab:last_N_layers}, we found that adding only the top N layers to the feature extractor will not result in huge performance drop. Such result suggests that the trainable parameter of CNN adapter can be further reduced in order to be more parameter efficient.
\subsubsection{Adapter layer shrinking}
To further explore the trade-off between parameter efficiency and performance, we shrink each adapter layer by decreasing its output channel. As shown in Table.\ref{tab:CNN shrink}, we can tell from the table that compressing the parameter of our CNN adapters inserted in the feature does not harm the performance very much, while the trainable parameters are  significantly eliminated. Such result shows that added CNN adapters can be even more parameter-efficient, which is worth exploring in the future.

\section{Conclusion}
\label{sec:conclusion}

In this paper, we propose CHAPTER, an efficient tuning technique specifically for SSL speech models. In addition, to insert adapter modules to the transformer layer, we also applied CNN adapters to the feature extractor to better adapt SSL speech models. Experiments were conducted to investigate the effectiveness of adapters on feature extractor; we found that this approach is very suitable for speaker-related tasks as well as emotion tasks. This opens a road for future work to investigate more on utilizing the feature extractor in SSL speech models in order to better capture speaker and emotion-related information.


\bibliographystyle{IEEEbib}
\bibliography{strings,refs}

\end{document}